\documentclass[acmsmall]{acmart}
\usepackage{cite,amsmath,amssymb,amsfonts,url,wrapfig,xspace,setspace,graphicx}
\usepackage[switch]{lineno}
\usepackage[T1]{fontenc}
\usepackage{hyperref}
\newcommand{\pct}[1]{#1\!\!\%\xspace}
\renewcommand{\c}{{\sf  C}\xspace}
\newcommand{\objc}{{\sf  Objective-C}\xspace}
\newcommand{\cpp}{{\sf  C++}\xspace}
\newcommand{\python}{{\sf  Python}\xspace}
\newcommand{\java}{{\sf  Java}\xspace}
\newcommand{\ruby}{{\sf  Ruby}\xspace}
\newcommand{\scala}{{\sf  Scala}\xspace}
\newcommand{\ts}{{\sf  TypeScript}\xspace}

\newcommand{\js}{{\sf  JavaScript}\xspace}
\newcommand{\haskell}{{\sf  Haskell}\xspace}
\newcommand{\cs}{{\sf  C\#}\xspace}
\newcommand{\go}{{\sf  Go}\xspace}
\newcommand{\coffee}{{\sf  CoffeeScript}\xspace}
\newcommand{\php}{{\sf  PHP}\xspace}
\newcommand{\perl}{{\sf  Perl}\xspace}
\newcommand{\clojure}{{\sf  Clojure}\xspace}
\newcommand{\erlang}{{\sf  Erlang}\xspace}
\newcommand{\smalltalk}{{\sf  SmallTalk}\xspace}
\newcommand{\gh}{{GitHub}\xspace}

\newcommand{\code}[1]{{\tt\small #1}\xspace}
\newcommand{\tabledesc}[1]{\begin{spacing}{.5}{\footnotesize #1}\end{spacing}}
\newcommand{\ea}{\emph{et al.}\xspace}    

\newcommand{\BOX}[1]{
\vspace{2.5mm}\noindent\fbox{\parbox{.97\columnwidth}{#1}}\vspace{1mm}}
\begin{document}
\newcommand{\numberOfProjectsWithDuplicates}{33\xspace}
\newcommand{\numDuplicateCommits}{27,450\xspace}
\newcommand{\percentageDuplicateCommits}{1.86\xspace}
\newcommand{\percentageDuplicateRowsLost}{5.28\xspace}
\newcommand{\smallProjectCommits}{26\xspace}
\newcommand{\percentageDuplicationReadjustmentRowsLost}{0.01\xspace}
\newcommand{\initialNumTSProjects}{41\xspace}
\newcommand{\initialNumTSCommits}{10,063\xspace}
\newcommand{\tsFirstCommit}{2,003-03-21\xspace}
\newcommand{\realTSProjNum}{16\xspace}
\newcommand{\realTSCommitsNum}{3,782\xspace}
\newcommand{\ratioOfTypeDefTSCommits}{34.6\xspace}
\newcommand{\tsValidRows}{2,475\xspace}
\newcommand{\tsValidCommits}{2,475\xspace}
\newcommand{\tsValidProjects}{13\xspace}
\newcommand{\percentageTypescriptRowsLost}{0.67\xspace}
\newcommand{\vCCommits}{16\xspace}
\newcommand{\vCppCommits}{7\xspace}
\newcommand{\vJavascriptCommits}{2,907\xspace}
\newcommand{\vPythonCommits}{488\xspace}
\newcommand{\percentageVEightRowsLost}{0.23\xspace}
\newcommand{\finalNumSha}{1,439,970\xspace}
\newcommand{\finalNumShaMio}{1.4\xspace}
\newcommand{\ratioReducedSha}{2.61\xspace}
\newcommand{\ratioReducedShaRows}{6.14\xspace}
\newcommand{\finalNumberOfProjectsIncluded}{708\xspace}
\newcommand{\finalSlocMio}{58.2\xspace}
\newcommand{\finalNumberAuthors}{46\xspace}
\newcommand{\finalNumberOfBugFixes}{517,770\xspace}

\newcommand{\MissingCommitsThousands}{106\xspace}
\newcommand{\MissingCommitsRatio}{19.95\xspace}
\newcommand{\PerlMissingRatio}{80\xspace}

\newcommand{\numberOfProjectsIncluded}{729\xspace}
\newcommand{\numberOfProjectsNotIncluded}{148\xspace}
\newcommand{\numberOfLargeProjectsNotIncluded}{101\xspace}
\newcommand{\slocMio}{80.7\xspace}
\newcommand{\numberCommitters}{29\xspace}
\newcommand{\numberAuthors}{47\xspace}
\newcommand{\linusCommitter}{73,038\xspace}
\newcommand{\linusAuthor}{11,343\xspace}
\newcommand{\numberCommitsMio}{1.5\xspace}
\newcommand{\numberOfBugFixes}{530,777\xspace}
\newcommand{\notInNewShaRaw}{6,440\xspace}
\newcommand{\notInEverythingRaw}{0\xspace}
\newcommand{\notInNewSha}{3,149\xspace}
\newcommand{\notInEverything}{1,387\xspace}
\newcommand{\notInNewShaWithProject}{3,156\xspace}
\newcommand{\notInEverythingWithProject}{54,424\xspace}

\newcommand{\ratioTestsFiles}{16.2\xspace}
\newcommand{\testFilesCommitted}{801,248\xspace}
\newcommand{\ratioTestFilesCommittedOverAll}{16.2\xspace}

\title{On the Impact of Programming Languages on Code Quality}
\subtitle{A Reproduction Study}
  
\author{Emery D. Berger}
\affiliation{\institution{University of Massachusetts Amherst}}

\author{Celeste Hollenbeck}
\affiliation{\institution{Northeastern University}}

\author{Petr Maj}
\affiliation{\institution{Czech Technical University in Prague}}

\author{Olga Vitek}
\affiliation{\institution{Northeastern University}}

\author{Jan Vitek}
\affiliation{\institution{Northeastern University} ~and \institution{Czech Technical University in Prague}}

\begin{abstract} 
This paper is a reproduction of work by Ray \ea which claimed to have
uncovered a statistically significant association between eleven programming
languages and software defects in projects hosted on \gh.  First we conduct
an experimental repetition, repetition is only partially successful, but it
does validate one of the key claims of the original work about the
association of ten programming languages with defects.  Next, we conduct a
complete, independent reanalysis of the data and statistical modeling steps
of the original study. We uncover a number of flaws that undermine the
conclusions of the original study as only four languages are found to have a
statistically significant association with defects, and even for those the
effect size is exceedingly small.  We conclude with some additional sources
of bias that should be investigated in follow up work and a few best
practice recommendations for similar efforts.
\end{abstract}
\maketitle

\renewcommand{\cite}{\citep}

\section{Introduction}

At heart, a programming language embodies a bet, the bet that a given set of
abstractions will increase developers' ability to deliver software that
meets its requirements.  Empirically quantifying the benefits of any set of
language features over others presents methodological challenges. While one
could have multiple teams of experienced programmers develop the same
application in different languages, such experiments are too costly to be
practical. Instead, when pressed to justify their choices, language
designers resort to intuitive arguments or proxies for productivity such as
numbers of lines of code.

However, large-scale, hosting services for code, such as \gh or
Source\-Forge, offer a glimpse into the life-cycles of software.  Not only
do they host the sources for millions of projects, but they also log changes
to their code. It is tempting to use these data to mine for broad patterns
across programming languages. The paper we reproduce here is an influential
attempt to develop a statistical model that relates various aspects of
programming language design to software quality.

\emph{What is the effect of programming language on software quality?} is
the question at the heart of the study by Ray \ea published at the 2014
Foundations of Software Engineering (FSE) conference~\cite{ray14}. The work
was sufficiently well-regarded in the software engineering community to be
nominated as a Communication of the ACM (CACM) \emph{research highlight}.
After another round of reviewing, a slightly edited version appeared
in journal form in 2017~\cite{ray17}. A subset of the authors also published
a short version of the work as a book chapter~\cite{book}.  The results
reported in the FSE paper and later repeated in the followup works are based
on an observational study of a corpus of 729 \gh projects written in 17
programming languages. To measure quality of code, the authors identified,
annotated, and tallied commits which were deemed to indicate bug fixes. The
authors then fit a Negative Binomial regression against the labeled data,
which was used to answer four research questions as follows:
\begin{enumerate}
\item[RQ1] ``{\bf Some languages have a greater association with defects
  than others}, although the effect is small.'' Languages associated with
  fewer bugs were \ts, \clojure, \haskell, \ruby, and \scala, while \c,
  \cpp, \objc, \js, \php and \python were associated with more bugs.
\item[RQ2] ``{\bf There is a small but significant relationship between
  language class and defects.} Functional languages have a smaller
  relationship to defects than either procedural or scripting languages.''
\item[RQ3] ``{\bf There is no general relationship between domain and language
  defect proneness.}'' Thus application domains are less important to
  software defects than languages.
\item[RQ4] ``{\bf Defect types are strongly associated with languages.} Some
  defect type like memory error, concurrency errors also depend on language
  primitives. Language matters more for specific categories than it does for
  defects overall.''
\end{enumerate}
Of these four results, it is the first two that garnered the most attention
both in print and on social media. This, likely, because they confirmed
commonly held beliefs about the benefits of static type systems and the
need to limit the use of side effects in programming.

\emph{Correlation is not causality, but it is tempting to confuse them.}
The original study couched its results in terms of \emph{associations} (i.e.
correlations) rather than \emph{effects} (i.e. \emph{causality}) and
carefully qualified effect size.  Unfortunately many of the paper's readers
were not as careful. The work was taken by many as a statement on the impact
of programming languages on defects. Thus one can find citations such as:
\begin{itemize}
\item ``\emph{...They found language design did have a significant, but
  modest effect on software quality.}''\,\cite{one}
\item ``\emph{...The results indicate that strong languages have better
  code quality than weak languages.}''\,\cite{three} 
\item ``\emph{...\c, \cpp and especially \objc is more failure prone than
  other languages.}''\,\cite{nine}
\end{itemize}

\begin{wraptable}{r}{4cm}\center{\footnotesize\sf
\vspace{-2mm}
\begin{tabular}{|r|r|r|}\hline
 & Cites & Self \\\hline
Cursory        &  77 & 1\\
Methods        &  12 & 0\\
Correlation    &   2 & 2\\
Causation      &  24 & 2\\\hline
\end{tabular}}
\caption{Citation analysis}\label{cites}
\vspace{-5mm}
\end{wraptable}

\noindent Table~\ref{cites} summarize our citation analysis. Of the 119
papers that were retrieved,\footnote{Retrieval performed on 12/01/18 based
  on the Google Scholar citations of the FSE paper, duplicates were
  removed.}  90 citations were either passing references (Cursory) or
discussed the methodology of the original study (Methods). Out of the
citations that discussed the results, 4 were careful to talk about
associations (i.e. correlation), while 26 used language that indicated
effects (i.e. causation). It is particularly interesting to observe that
even the original authors, when they cite their own work sometimes resort to
causal language: Ray and Posnett write ``\emph{Based on our previous study
  \cite{ray14} we found that the overall effect of language on code quality
  is rather modest.}''~\cite{book}, Devanbu writes ``\emph{We found that
  static typing is somewhat better than dynamic typing, strong typing is
  better than weak typing, and built-in memory management is
  better}''~\cite{prem}, and ``\emph{Ray [...] said in an interview that
  functional languages were boosted by their reliance on being mathematical
  and the likelihood that more experienced programmers use
  them.}''~\cite{info}.  Section 2 of the present paper gives a detailed
account of the original study and its conclusions.

Given the controversy generated by the CACM paper on social media, and some
surprising observations in the text of the original study (\emph{e.g.}, that
V8 is their largest \js project -- when the virtual machine is written in
\cpp), we wanted to gain a better understanding of the exact nature of the
scientific claims made in the study and how broadly they are actually
applicable. To this end, we chose to conduct an independent reproduction
study.

\begin{wrapfigure}{r}{6cm}\begin{center}
\vspace{-2mm}
\includegraphics[width=.4\columnwidth]{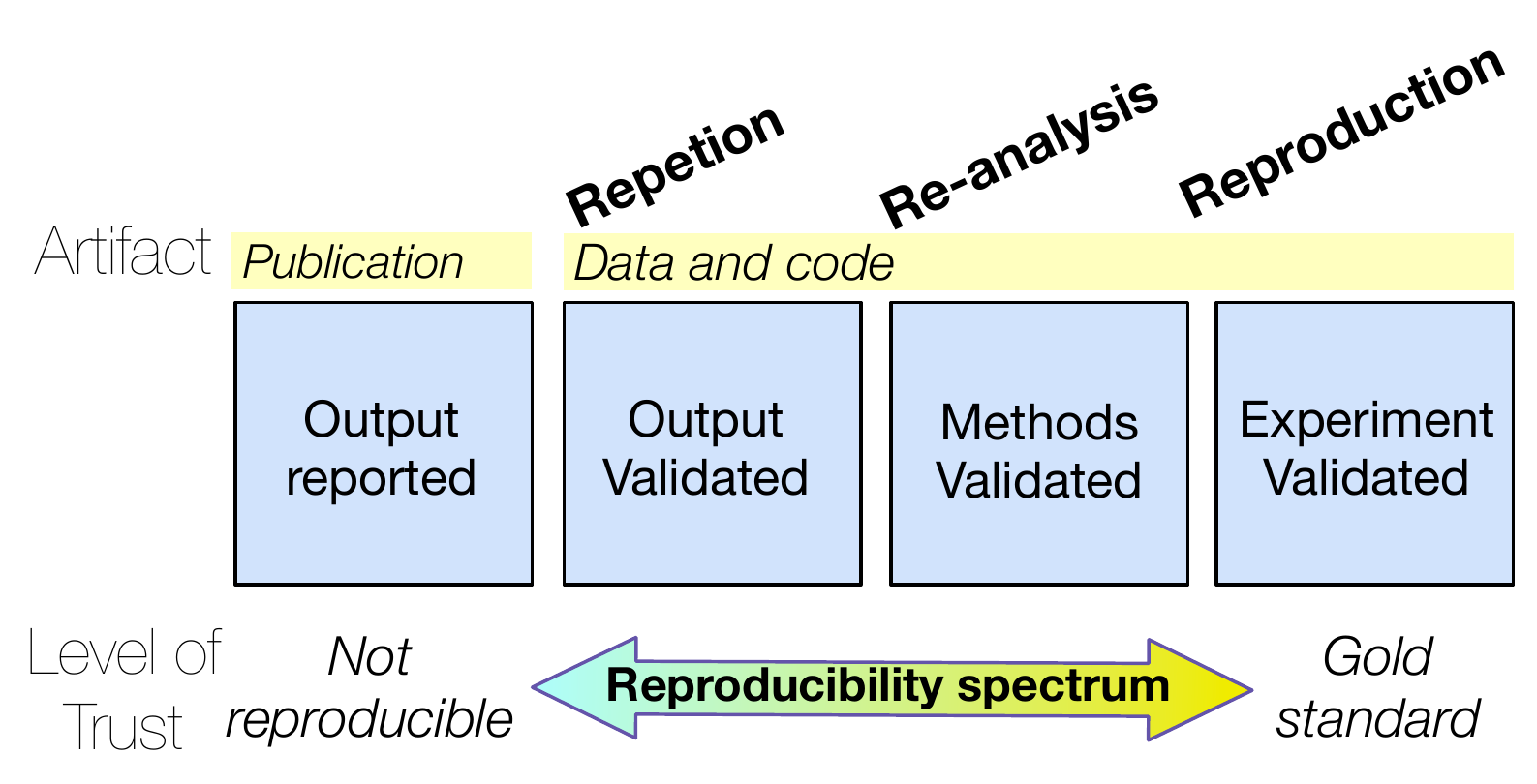}\end{center}
\caption{Reproducibility spectrum (from \citep{spectrum})}\label{spec}
\vspace{-3mm}
\end{wrapfigure}

\paragraph{Methodology}
Reproducibility of results is not a binary proposition. Instead, it is a
spectrum of objectives that provide assurances of different kinds (see
Figure~\ref{spec} using terms from~\cite{terms,emsoft11}).

\noindent \emph{Experimental repetition} aims to replicate the results of
some previous work with the same data and methods and should yield the same
numeric results.  Repetition is the basic guarantee provided by artifact
evaluation~\cite{cacm15}.  \emph{Reanalysis} examines the robustness of the
conclusions to the methodological choices. Multiple analysis methods may be
appropriate for a given dataset, and the conclusions should be robust to the
choice of method. Occasionally, small errors may need to be fixed, but the
broad conclusions should hold. Finally, \emph{Reproduction} is the gold
standard; it implies a full-fledged independent experiment conducted with
different data and the same or different methods. To avoid bias, repetition,
reanalysis, and reproduction are conducted independently.  The only contact
expected with the original authors is to request their data and code.

\paragraph{Results} We began with an experimental repetition, conducting
it in a similar fashion to a conference artifact evaluation~\cite{cacm15}
(Section 3 of the paper).  The repetition was partially successful. We were
\begin{wrapfigure}{r}{3cm}\begin{center}\vspace{-3mm}
\includegraphics[width=.2\columnwidth]{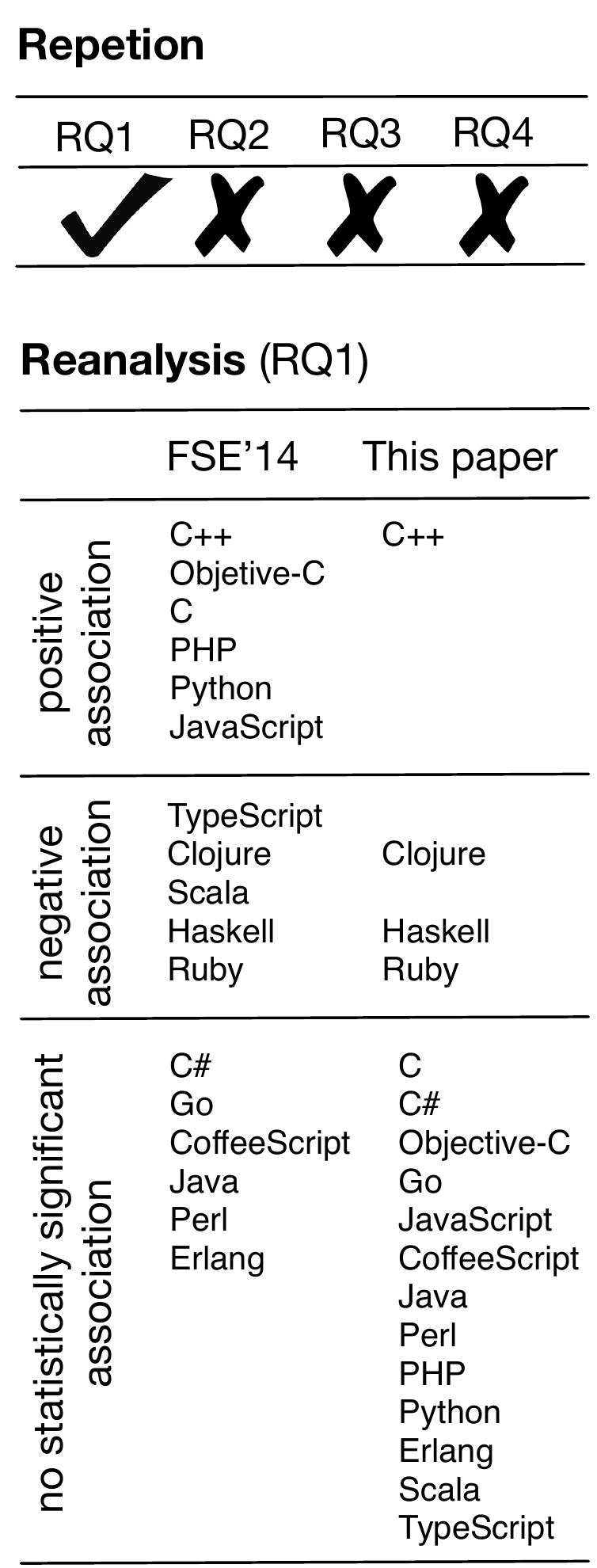}\end{center}
\caption{Result summary}\label{sum}
\vspace{-5mm}
\end{wrapfigure}
\noindent  
able to mostly replicate RQ1 based on the artifact provided by the
authors. We found ten languages with a statistically significant association
with errors, instead of the eleven reported. For RQ2, we uncovered
classification errors that made our results depart from the published
ones. Lastly, RQ3 and RQ4 could not be repeated due to missing code and
discrepancies in the data.

For \emph{reanalysis}, we focused on RQ1 and discovered substantial issues
(Section 4 of this paper).  While the original study found that 11 out of 17
languages were correlated with a higher or lower number of defective
commits, upon cleaning and reanalyzing the data, the number of languages
dropped to 7. Investigations of the original statistical modeling revealed
technical oversights such as inappropriate handling of multiple hypothesis
testing. Finally, we enlisted the help of independent developers to
cross-check the original method of labeling defective commits, this led us
to estimate a false positive rate of 36\% on buggy commit labels. Combining
corrections for all of these aforementioned items, the reanalysis revealed
that only 4 out of the original 11 languages correlated with abnormal defect
rates, and even for those the effect size is exceedingly small.

Figure~\ref{sum} summarizes our results: Not only is it not possible to
establish a causal link between programming language and code quality based
on the data at hand, but even their correlation proves questionable.  Our
analysis is repeatable and available in an artifact hosted at:
{\small\url{https://tinyurl.com/ybrwo4mj}}.

\paragraph{Follow up work}
While reanalysis was not able to validate the results of the original study,
we stopped short of conducting a reproduction as it is unclear what that
would yield.  In fact, even if we were to obtain clean data and use the
proper statistical methods, more research is needed to understand all the
various sources of bias that may affect the outcomes.  Section~\ref{threats}
lists some challenges that we discovered while doing our repetition. For
instance, the ages of the projects vary across languages (older languages
such as C are dominated by mature projects such as Linux), or the data
includes substantial numbers of commits to test files (how bugs in tests are
affected by language characteristics is an interesting question for future
research).  We believe that there is a need for future research on this
topic, we thus conclude our paper with some best practice recommendations
for future researchers (Section~\ref{best}).

\section{Original Study and Its Conclusions}

\subsection{Overview}

The FSE paper by Ray \ea \cite{ray14} aimed to explore associations between
languages, paradigms, application domains, and software defects from a
real-world ecosystem across multiple years. Its multi-step, mixed-method
approach included collecting commit information from \gh; identifying each
commit associated with a bug correction; and using Negative Binomial
Regression (NBR) to analyze the prevalence of bugs.  The paper answers
claims answer the following questions.

\BOX{RQ1. \it Are some languages more defect prone than others?}

The paper concluded that ``\emph{Some languages have a greater association
  with defects than others, although the effect is small}.''  Results appear
in a table that fits an NBR model to the data; it reports coefficient
estimates, their standard errors, and ranges of p-values.  The authors noted
that confounders other than languages explained most of the variation in the
number of bug-fixing commits, quantified by analysis of deviance. They
reported p-values below .05, .01, and .001 as ``statistically significant''.
Based on these associations, readers may be tempted to conclude that \ts,
\haskell, \clojure, \ruby, and \scala were less error-prone; and \cpp,
\objc, \c, \js, \php, and \python were more error-prone.

\BOX{RQ2. \it Which language properties relate to defects?}

The study concluded that ``\emph{There is a small but significant
  relationship between language class and defects. Functional languages have
  a smaller relationship to defects than either procedural or scripting
  languages.}''  The impact of nine language categories across four classes
was assessed. Since the categories were highly correlated (and thus
compromised the stability of the NBR), the paper modeled aggregations of the
languages by class. The regression included the same confounders as in RQ1
and represented language classes. The authors report the coefficients, their
standard errors, and ranges of p-values. Results may lead readers to
conclude that functional, strongly typed languages induced fewer errors,
while procedural, weakly typed, unmanaged languages induced more errors.

\BOX{RQ3. \it Does language defect proneness depend on domain?}

The study used a mix of automatic and manual methods to classify projects
into six application domains. After removing outliers, and calculating the
Spearman correlation between the order of languages by bug ratio within
domains against the order of languages by bug ratio for all domains, it
concluded that ``\emph{There is no general relationship between domain and
  language defect proneness}''.  The paper states that all domains show
significant positive correlation, except the Database domain.  From this,
readers might conclude that the variation in defect proneness comes from the
languages themselves, making domain a less indicative factor.

\BOX{RQ4. \it What's the relation between language \& bug category?}

The study concluded that ``\emph{Defect types are strongly associated with
  languages; Some defect type like memory error, concurrency errors also
  depend on language primitives. Language matters more for specific
  categories than it does for defects overall.}'' The authors report that
88\% of the errors fall under the general Programming category, for which
results are similar to RQ1.  Memory Errors account for 5\% of the bugs,
Concurrency for 2\%, and Security and other impact errors for 7\%. For
Memory, languages with manual memory management have more errors. \java
stands out; it is the only garbage collected language associated with more
memory errors. For Concurrency, inherently single-threaded languages
(\python, \js, etc.) have fewer errors than languages with concurrency
primitives.  The causal relation for Memory and Concurrency is
understandable, as the classes of errors require particular language
features.

\subsection{Methods in the original study}

Below, we summarize the process of data analysis by the original manuscript
while splitting it into phases: data acquisition, cleaning, and modeling.

\renewcommand{\SS}{\hspace{-.7mm}} 

\subsubsection{Data Acquisition}

For each of the 17 languages with the most projects on \gh, 50 projects with
the highest star rankings were selected.  Any project with fewer than 28
commits was filtered out, leaving 729 projects (86\%).  For each project,
commit histories were collected with {\tt git log --no-merges
  --numstat}. The data were split into rows, such that each row had a unique
combination of file name, project name, and commit identifier.  Other fields
included committer and author name, date of the commit, commit message, and
number of lines inserted and deleted.  In summary, the original paper states
that the input consisted of {729} projects written in {17} languages,
accounting for {63} million SLOC created over {1.5} million commits written
by {29} thousand authors. Of these, {566,000} commits were bug fixes.
  
\subsubsection{Data Cleaning}

As any project may have code written in multiple languages, each row of the
data was labeled by language based on the file's extension (\ts is
.\code{ts}, and so on).  To rule out small projects, projects with fewer
than 20 commits in any single language were filtered out for that
language. The next step labeled commits as bug fixes, adding language
classes to each commit. Bug fixes were found by searching for error-related
keywords: \emph{error, bug, fix, issue, mistake, incorrect, fault, defect}
and \emph{flaw} in commit messages, similar to a heuristic found in Mockus
and Votta~\cite{Mockus00}.  Each row of the data was labeled with four
classes. A {Paradigm} was one of procedural, functional, or scripting. The
{Compile} class indicated whether a language was statically or dynamically
typed. The {Type} class indicated whether a language admitted
`type-confusion', i.e., it allowed interpreting a memory region populated by
a value of one type as another type. A language was strongly typed if it
explicitly detected type confusion and reported it as such. The {Memory}
class indicated whether the language required developers to manage memory by
hand.

\subsubsection{Statistical Modeling}

For RQ1, the manuscript specified an NBR~\cite{faraway}, where an
observation was a combination of project and language. In other words, a
project written in three languages had three observations.  For each
observation, the regression used bug-fixing commits as a response variable,
and the languages as the independent variables.  NBR is an appropriate
choice, given the non-negative and discrete nature of the counts of
commits. To adjust for differences between the observations, the regression
included the confounders age, number of commits, number of developers, and
size (represented by inserted lines in commits), all log-transformed to
improve the quality of fit. For the purposes of RQ1, the model for an
observation $i$ was:

\vspace{-3mm}
{\footnotesize\begin{eqnarray}
&\mathrm{bcommits}_i\sim\mathit{NegativeBinomial}(\mu_i,\theta),\ \mbox{where}&\nonumber\\
&\ \mathrm{E}\{\mathrm{bcommits}_i\}=\mu_i&\nonumber\\ 
&\mathrm{Var}\{\mathrm{bcommits}_i\}=\mu_i+\mu_i^2/\theta& \nonumber\\
& log\,\mu_i=\beta_0+\beta_1 \mathrm{log(commits)}_i+\beta_2 \mathrm{log(age)}_i+ \beta_3 \mathrm{log(size)}_i+\beta_4 \mathrm{log(devs)}_i+\sum_{j=1}^{16}\beta_{(4+j)}\mathrm{language}_{ij} & \nonumber 
\end{eqnarray}}
\noindent The programming languages were coded with {\it weighted
  contrasts}. These contrasts were customized in a way to interpret
$\beta_0$ as the average log-expected number of bugs in the {\it dataset}.
Therefore, $\beta_5,\ldots,\beta_{20}$ were the deviations of the
log-expected number of bug-fixing commits in a language from the average of
the log-expected number of bug-fixing commits {\it in the dataset}. Finally,
the coefficient $\beta_{21}$ (corresponding to the last language in
alphanumeric order) was derived from the contrasts after the model
fit~\cite{knnl}. Coefficients with a statistically significant negative
value indicated a lower expected number of bug-fixing commits; coefficients
with a significant positive value indicated a higher expected number of
bug-fixing commits.  The model-based inference of parameters
$\beta_5,\ldots,\beta_{21}$ was the main focus of RQ1.

For RQ2, the study fit another NBR, with the same confounder variables, to
study the association between language classes and the number of bug-fixing
commits.  It then used Analysis of Deviance to quantify the variation
attributed to language classes and the confounders. For RQ3, the paper
calculated the Spearman's correlation coefficient between defectiveness by
domain and defectiveness overall, with respect to language, to discuss the
association between languages versus that by domain. For RQ4, the study once
again used NBR, with the same confounders, to explore propensity for
bugfixes among the languages with regard to bug types. 

\section{Experimental Repetition}

Our first objective is to repeat the analyses of the FSE paper and to
obtain the same results. We requested and received from the original authors
an artifact containing 3.45~GB of processed data and 696~lines of R code to
load the data and perform statistical modeling steps.

\subsection{Methods}

Ideally, a repetition should be a simple process, where a script generates
results and these match the results in the published paper. In our case, we
only had part of the code needed to generate the expected tables and no code
for graphs.  We therefore wrote new R scripts to mimic all of the steps, as
described in the original manuscript.  We found it essential to automate the
productions of all numbers, tables, and graphs shown in our paper as we had
to iterate multiple times. The code for repetition amounts to 1,140~lines of
R (file \code{\small repetition.Rmd} and \code{\small implementation.R} in
our artifact).

\subsection{Results}

The data was given to us in the form of two CSV files. The first, larger
file contained one row per file and commit, and it contained the bug fix
labels.  The second, smaller file aggregated rows with the same commit and
the same language. Upon preliminary inspection, the files contained
information on \numberOfProjectsIncluded projects and \numberCommitsMio
million commits.  We found an additional \numberOfProjectsNotIncluded
projects that were not omitted from the original study without explanation.

\paragraph{Developers vs. Committers}
One discrepancy was the \numberAuthors thousand authors we observed versus
the 29 thousand reported. This is explained by the fact that, although the
FSE paper claimed to use \emph{developers} as a control variable, it was in
fact counting \emph{committers}: a subset of developers with commit
rights. For instance, Linus Torvalds has \linusCommitter commits, of which
he personally authored \linusAuthor, the remaining are due to other members
of the project.  The rationale for using developers as a control variable is
that the same individual may be more or less prone to committing bugs, but
this argument does not hold for committers as they aggregate the work of
multiple developers. We chose to retain committers for our reproduction but
note that this choice should be revisited in follow up work.

\begin{table*}[b!]
\centering
\caption{Negative Binomial Regression for Languages}
\begingroup\small
\scalebox{0.9}{
\begin{tabular}{@{}r||ll|ll||ll@{}}
  \hline
  \rule{0pt}{3ex} & \multicolumn{4}{c||}{\normalsize Original Authors}  & \multicolumn{2}{c}{\normalsize Repetition}\\[1mm] & \multicolumn{2}{c|}{(a) FSE~\cite{ray14}} & \multicolumn{2}{c||}{(b) CACM~\cite{ray17}} & \multicolumn{2}{c}{(c)}\\ & Coef & P-val & Coef & P-val & Coef & P-val\\ \hline
Intercept & -1.93 & <0.001 & \cellcolor{white} -2.04 & \cellcolor{white} <0.001 & \cellcolor{white} -1.8 & \cellcolor{white} <0.001 \\ 
  log commits & ~2.26 & <0.001 & \cellcolor{white} ~0.96 & \cellcolor{white} <0.001 & \cellcolor{white} ~0.97 & \cellcolor{white} <0.001 \\ 
  log age & ~0.11 & <0.01 & \cellcolor{white} ~0.06 & \cellcolor{white} <0.001 & \cellcolor{white} ~0.03 & \cellcolor{white} \hphantom{<}0.03 \\ 
  log size & ~0.05 & <0.05 & \cellcolor{white} ~0.04 & \cellcolor{white} <0.001 & \cellcolor{white} ~0.02 & \cellcolor{white} <0.05 \\ 
  log devs & ~0.16 & <0.001 & \cellcolor{white} ~0.06 & \cellcolor{white} <0.001 & \cellcolor{white} ~0.07 & \cellcolor{white} <0.001 \\ 
   \hline
\hline
C & ~0.15 & <0.001 & \cellcolor{white} ~0.11 & \cellcolor{white} <0.01 & \cellcolor{white} ~0.16 & \cellcolor{white} <0.001 \\ 
  C++ & ~0.23 & <0.001 & \cellcolor{white} ~0.18 & \cellcolor{white} <0.001 & \cellcolor{white} ~0.22 & \cellcolor{white} <0.001 \\ 
  C\# & ~0.03 & -- & \cellcolor{white} -0.02 & \cellcolor{white} -- & \cellcolor{white} ~0.03 & \cellcolor{white} \hphantom{<}0.602 \\ 
  Objective-C & ~0.18 & <0.001 & \cellcolor{white} ~0.15 & \cellcolor{white} <0.01 & \cellcolor{white} ~0.17 & \cellcolor{white} \hphantom{<}0.001 \\ 
  Go & -0.08 & -- & \cellcolor{white} -0.11 & \cellcolor{white} -- & \cellcolor{white} -0.11 & \cellcolor{white} \hphantom{<}0.086 \\ 
  Java & -0.01 & -- & \cellcolor{white} -0.06 & \cellcolor{white} -- & \cellcolor{white} -0.02 & \cellcolor{white} \hphantom{<}0.61 \\ 
  Coffeescript & -0.07 & -- & \cellcolor{white} ~0.06 & \cellcolor{white} -- & \cellcolor{white} ~0.05 & \cellcolor{white} \hphantom{<}0.325 \\ 
  Javascript & ~0.06 & <0.01 & \cellcolor{gray!25} ~0.03 & \cellcolor{gray!25} -- & \cellcolor{white} ~0.07 & \cellcolor{white} <0.01 \\ 
  Typescript & -0.43 & <0.001 & \cellcolor{gray!25} ~0.15 & \cellcolor{gray!25} -- & \cellcolor{white} -0.41 & \cellcolor{white} <0.001 \\ 
  Ruby & -0.15 & <0.05 & \cellcolor{white} -0.13 & \cellcolor{white} <0.01 & \cellcolor{white} -0.13 & \cellcolor{white} <0.05 \\ 
  Php & ~0.15 & <0.001 & \cellcolor{white} ~0.1 & \cellcolor{white} <0.05 & \cellcolor{gray!25} ~0.13 & \cellcolor{gray!25} \hphantom{<}0.009 \\ 
  Python & ~0.1 & <0.01 & \cellcolor{white} ~0.08 & \cellcolor{white} <0.05 & \cellcolor{white} ~0.1 & \cellcolor{white} <0.01 \\ 
  Perl & -0.15 & -- & \cellcolor{white} -0.12 & \cellcolor{white} -- & \cellcolor{white} -0.11 & \cellcolor{white} \hphantom{<}0.218 \\ 
  Clojure & -0.29 & <0.001 & \cellcolor{white} -0.3 & \cellcolor{white} <0.001 & \cellcolor{white} -0.31 & \cellcolor{white} <0.001 \\ 
  Erlang & ~0 & -- & \cellcolor{white} -0.03 & \cellcolor{white} -- & \cellcolor{white} ~0 & \cellcolor{white} \hphantom{<}1 \\ 
  Haskell & -0.23 & <0.001 & \cellcolor{white} -0.26 & \cellcolor{white} <0.001 & \cellcolor{white} -0.24 & \cellcolor{white} <0.001 \\ 
  Scala & -0.28 & <0.001 & \cellcolor{white} -0.24 & \cellcolor{white} <0.001 & \cellcolor{white} -0.22 & \cellcolor{white} <0.001 \\ 
   \hline
\end{tabular}
}
\endgroup
\label{langtable}
\end{table*}

\paragraph{Measuring code size}
The commits represented \slocMio million lines of code. We could not account
for a difference of 17 million SLOC from the reported size.  We also remark,
but do not act on, the fact that project size, computed in the FSE paper as
the sum of inserted lines, is not accurate---as it does not take deletions
into account.  We tried to subtract deleted lines and obtained projects with
negative line counts. This is due to the treatments of Git merges. A merge
is a commit which combines conflicting changes of two parent commits.  Merge
commits are not present in our data; only parent commits are used, as they
have more meaningful messages.  If both parent commits of a merge delete the
same lines, the deletions are double counted.  It is unclear what the right
metric of size should be.

\subsubsection{Are some languages more defect prone than others (RQ1)}

We were able to qualitatively (although not exactly) repeat the result of
RQ1. Table~\ref{langtable} (a) has the original results, and (c) has our
repetition. One disagreement in our repetition is with \php. The FSE paper
reported a p-value <$.001$, while we observed <$.01$, in other words, \php
the association of \php with defects is not statistically significant. The
original authors corrected that value in their CACM repetition (shown in
Table~\ref{langtable}(b)), so this may just be a reporting error. On the
other hand, the CACM paper dropped the significance of \js and \ts without
explanation. The other difference is in the coefficients for the control
variables. Upon inspection of the code, we noticed that the original
manuscript used a combination of log and log10 transformations of these
variables, while the repetition consistently used log. The CACM repetition
fixed this problem.

\subsubsection{Which language properties relate to defects (RQ2)}

As we approached RQ2, we faced an issue with the language categorization
used in the FSE paper. The original categorization is reprinted in
Table~\ref{classestable}.  The intuition is that each category should group
languages that have ``similar'' characteristics along some axis of language
design.

\begin{table}[!t]
\center
\caption{Language classes defined by the FSE paper.}\label{classestable}
{\small\begin{tabular}{l@{~}lp{7.7cm}}\hline
Classes & Categories & Languages\\\hline
\bf Paradigm & Procedural &
{\c\SS} {\cpp\SS} {\cs\SS} {\objc\SS} {\java\SS} {\go\SS}\\
& Scripting &
{\coffee\SS} {\js\SS} {\python\SS}  {\perl\SS} {\php\SS} {\ruby\SS}\\
& Functional & {\clojure\SS} {\erlang\SS} {\haskell\SS} {\scala\SS}\\\hline
\bf Compilation & Static &
{\c\SS} {\cpp\SS} {\cs\SS} {\objc\SS} {\java\SS} {\go\SS}  {\haskell\SS} {\scala\SS} \\
&Dynamic& {\coffee\SS} {\js\SS} {\python\SS} {\perl\SS} {\php\SS} {\ruby\SS} 
{\clojure\SS} {\erlang\SS}\\\hline
\bf Type & Strong&
{\cs\SS} {\java\SS} {\go\SS} {\python\SS} {\ruby\SS} {\clojure\SS} {\erlang\SS} {\haskell\SS} {\scala\SS}\\
&Weak&
{\c\SS} {\cpp\SS} {\objc\SS} {\php\SS} {\perl\SS} {\coffee\SS} {\js\SS} \\\hline
 \bf Memory & Unmanaged & {\c\SS} {\cpp\SS} {\objc\SS}\\
& Managed & Others\\\hline
\end{tabular}}
\end{table}


\begin{table}[b!]\center
\caption{Negative Binomial Regression for Language Classes}
\begin{tabular}{@{}r||ll||ll|ll@{}}  \hline
\rule{0pt}{3ex} & \multicolumn{2}{c||}{\normalsize (a) Original} & \multicolumn{2}{c}{\normalsize (b) Repetition} & \multicolumn{2}{c}{\normalsize (c) Reclassification}\\[1mm] & Coef & P-val & Coef & P-val & Coef & P-val \\ \hline
Intercept           & -2.13 & <0.001 &-2.14 &  <0.001 &  -1.85 &  <0.001 \\ 
            log age & ~0.07 & <0.001 &~0.15 &  <0.001 &  ~0.05 &  0.003 \\ 
           log size & ~0.05 & <0.001 &~0.05 &  <0.001 &  ~0.01 &  0.552 \\ 
           log devs & ~0.07 & <0.001 &~0.15 &  <0.001 &  ~0.07 &  <0.001 \\ 
        log commits & ~0.96 & <0.001 &~2.19 &  <0.001 &  ~1    &  <0.001 \\    \hline\hline
\tt Fun Sta Str Man & -0.25 & <0.001 &-0.25 &  <0.001 &  -0.27 &  <0.001 \\ 
\tt Pro Sta Str Man & -0.06 &  <0.05 &-0.06 &   0.039 & \cellcolor{gray!25} -0.03 & \cellcolor{gray!25} 0.24 \\ 
\tt Pro Sta Wea Unm & ~0.14 & <0.001 &~0.14 &  <0.001 &  ~0.19 &  0 \\ 
\tt Scr Dyn Wea Man & ~0.04 &  <0.05 &~0.04 &  0.018  &\cellcolor{gray!25} ~0 & \cellcolor{gray!25} 0.86 \\ 
\tt Fun Dyn Str Man & -0.17 & <0.001 &-0.17 &  <0.001 &   -- &  -- \\ 
\tt Scr Dyn Str Man &~0.001 &     -- &   ~0 &   0.906 &   -- &  -- \\ 
\tt Fun Dyn Wea Man &    -- &     -- &   -- &      -- & -0.18 &  <0.001 \\  \hline
\end{tabular}\label{t5}
\medskip \tabledesc{Language classes are combined procedural (Pro),
  functional (Fun), scripting (Scr), dynamic (Dyn), static (Sta), strong
  (Str), weak (Wea), managed (Man), and unmanaged (Unm). Rows marked -- have
  no observation.}
\end{table}

The first thing to observe is that any such categorization will have some
unclear fits. The original authors admitted as much by excluding \ts from
this table, as it was not obvious whether a gradually typed language is
static or dynamic.

But there were other odd ducks. \scala is categorized as a functional
language, yet it allows programs to be written in an imperative manner. We
are not aware of any study that shows that the majority of \scala users
write functional code. Our experience with \scala is that users freely mix
functional and imperative programming. \objc is listed as a statically
compiled and unmanaged language. Yet it has an object system that is
inspired by \smalltalk, thus its treatment of objects is quite dynamic, and
objects are collected by reference counting, thus memory is partially
managed.

The Type category is the most counter-intuitive for programming language
experts as it expresses whether a language allows value of one type to be
interpreted as another, e.g. due to automatic conversion. The CACM paper
attempted to clarify this definition with the example of the \objc \code{ID}
type which can hold any value.  If this is what the authors intend, then
\python, \ruby, \clojure, and \erlang would be weak as they have similar 
generic types.

In our repetition, we modified the categories accordingly and introduced a
new category of {Functional-Dynamic-Weak-Managed} to accommodate \clojure
and \erlang. Table~\ref{t5}(c) summarizes the results with the new
categorization. The reclassification (using zero-sum contrasts introduced in
section~\ref{sec:zerosum}) disagrees on the significance of 2 out of 5
categories.

\subsubsection{Does language defect proneness depend on domain (RQ3)}

We were unable to repeat RQ3, as the artifact did not include code to
compute the results. However, the data contained the classification of
projects in domains, which allowed us to attempt to recreate part of the
analysis described in the paper.  While we successfully replicated the
initial analysis step, we could not match the removal of outliers described
in the FSE paper.  Stepping outside of the repetition, we explore an
alternative approach to answer the question.  Table~\ref{domainNBR} uses an
NBR with domains instead of languages. The results suggest there is no
evidence that the application domain is a predictor of bug-fixes as the
paper claims.

\begin{table}[!h]\centering \caption{NBR for RQ3}
\begin{tabular}{@{}r|rl@{}}  \hline
  \rule{0pt}{3ex} & Coef & p-Val \\ \hline \hline
(Intercept) & -1.94 & <0.001 \\ 
  log age & ~0.05 & <0.001 \\ 
  log size & ~0.03 & <0.001 \\ 
  log devs & ~0.08 & <0.001 \\ 
  log commits & ~0.96 &<0.001 \\  \hline \hline
Application & ~0 & 1.00 \\ 
  CodeAnalyzer & -0.05 & 0.93 \\ 
  Database & ~0.04 & 1.00 \\ 
  Framework & ~0.01 & 1.00 \\ 
  Library & -0.06 & 0.23 \\ 
  Middleware & ~0 & 1.00 \\    \hline
\end{tabular}\label{domainNBR}
\end{table}

\subsubsection{ What's the relation between language \& bug category (RQ4)}

We were unable to repeat the results of RQ4 because the artifact did not
contain the code which implemented the heatmap or NBR for bug
types. Additionally, we found no single column in the data that contained
the bug categories reported in the FSE paper. It was further unclear whether
the bug types were disjoint: adding together all of the percentages for
every bug type mentioned in Table~5 of the FSE study totaled 104\%.  The
input CSV file did contain two columns which, combined, matched these
categories. When we attempted to reconstruct the categories and compared
counts of each bug type, we found discrepancies with those originally
reported.  For example, we had 9 times as many Unknown bugs as the original,
but we had only less than half Memory bugs.  Such discrepancies make
repetition invalid.

\subsection{Outcome}

The repetition was partly successful. RQ1 produced small differences, but
qualitatively similar conclusions. RQ2 could be repeated, but we noted
issues with language classification; fixing these issues changed the outcome
for 2 out of 5 categories.  RQ3 could not be repeated, as the code was
missing and our reverse engineering attempts failed. RQ4 could not be
repeated due to irreconcilable differences in the data. 

\section{Reanalysis}

Our second objective is to carry out a reanalysis of RQ1 of the FSE paper.
The reanalysis differs from repetition in that it proposes alternative data
processing and statistical analyses to address what we identify as
methodological weaknesses of the original work.

\subsection{Methods: Data processing}\label{sec:methods:dataprocess}

First, we examined more closely the process of data acquisition in the
original work. This step was intended as a quality control, and it did not
result in changes to the data. 

We wrote software to automatically download and check commits of projects
against \gh histories. Out of 729 projects used in the FSE paper, 618 could
be downloaded. The other projects may have been deleted or became private. 

The downloaded projects were matched by name. As the FSE data lacked project
owner names, the matches were ambiguous.  By checking for matching SHAs, we
confidently identified 423 projects as belonging to the study.  

For each matched project, we compared its entire history of commits to its
commits in the FSE dataset, as follows. We identified the most recent commit
$c$ occurring in both. Commits chronologically older than $c$ were
classified as either {\it valid} (appearing in the original study), {\it
  irrelevant} (not affecting language files), or {\it missing} (not
appearing in the original study).  

\begin{figure}[!h]
\includegraphics[width=.9\columnwidth]{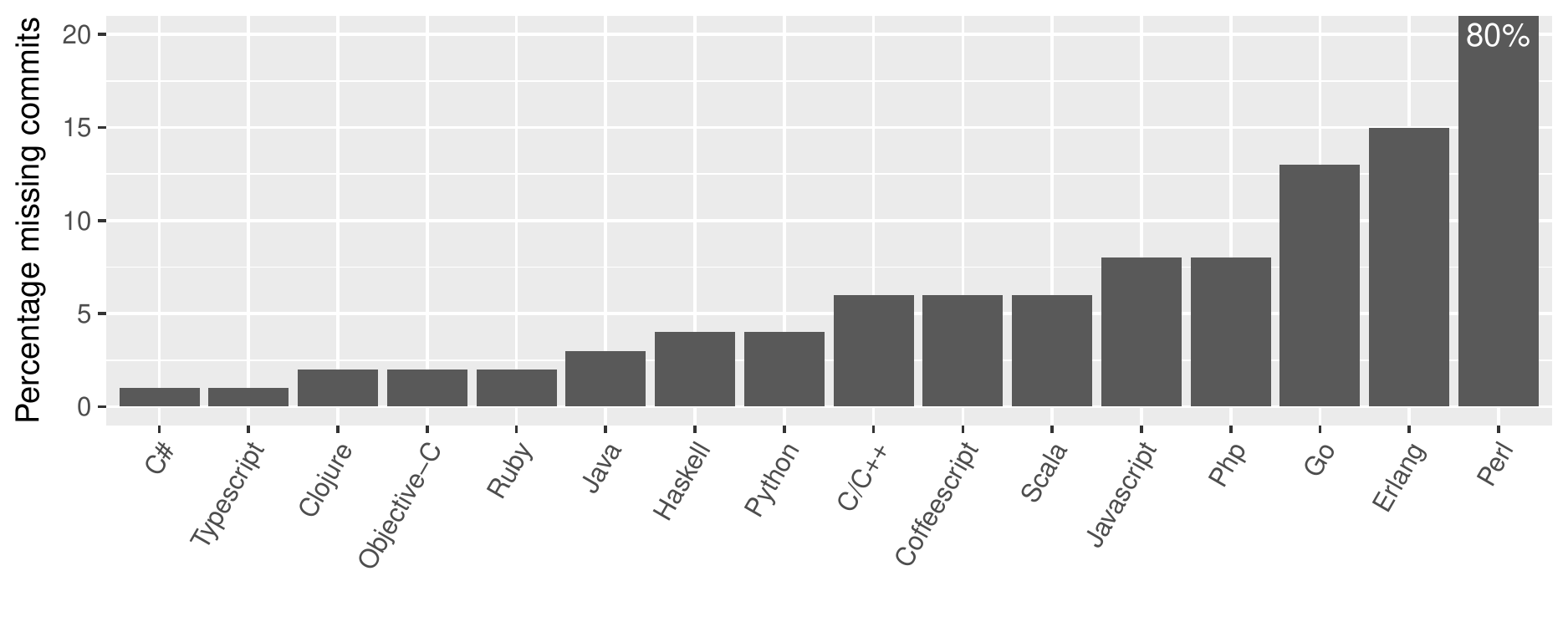}
\vspace{-8mm}
\caption{Percentage of commits identified as missing from the FSE dataset.}\label{miss}
\end{figure}

We found \MissingCommitsThousands~K missing commits
(i.e. \pct{\MissingCommitsRatio} of the dataset). Perl stands out with
\pct{\PerlMissingRatio} of commits that were missing in the original
manuscript (Fig.~\ref{miss} lists the ratio of missing commits per
language). Manual inspection of a random sample of the missing commits did
not reveal any pattern.  

We also recorded {\it invalid} commits (occurring in the study but absent
from the \gh history). Four projects had substantial numbers of invalid
commits, likely due to matching errors or a change in commit history (such
as with the \code{git rebase} command).

Next, we applied data cleaning steps (see below for details), each of these
was necessary to compensate for errors in data acquisition of the original
study:
\begin{enumerate}
\item \emph{Deduplication}
\item \emph{Removal of TypeScript}
\item \emph{Accounting for C and C++}
\end{enumerate}
Our implementation consists of 1323 lines of R code split between files
\code{\small re-analysis.Rmd} and \code{\small implem\-entation.R} in the
artifact.

\subsubsection{Deduplication}
While the input data did not include forks, we checked for project
similarities by searching for projects with similar commit identifiers.  We
found \numberOfProjectsWithDuplicates projects that shared one or more
commits. Of those, 18 were related to \code{bitcoin}, a popular project that
was frequently copied and modified. The projects with duplicate commits are:
{\newcommand{\xxx}[1]{{\sf\small #1}}
\xxx{litecoin},
\xxx{mega}-\xxx{coin},
\xxx{memorycoin},
\xxx{bitcoin},
\xxx{bitcoin-qt-i2p},
\xxx{anoncoin},
\xxx{smallchange},
\xxx{primecoin},
\xxx{terracoin},
\xxx{zetacoin},
\xxx{datacoin},
\xxx{datacoin-hp},
\xxx{freicoin},
\xxx{ppcoin},
\xxx{namecoin},
\xxx{namecoin-qt},
\xxx{namecoinq},
\xxx{ProtoShares},
\xxx{QGIS},
\xxx{Quantum-GIS},
\xxx{incub\-ator-spark},
\xxx{spark},          
\xxx{sbt},
\xxx{xsbt},
\xxx{Play20},
\xxx{playframework},
\xxx{ravendb},
\xxx{SignalR},
\xxx{Newtonsoft.Json},
\xxx{Hystrix},
\xxx{RxJava},
\xxx{clojure-scheme},
\xxx{clojurescript}}.
In total, there were \numDuplicateCommits duplicated commits, or
\pct{\percentageDuplicateCommits} of all commits. We deleted these commits
from our dataset to avoid double counting some bugs.

\subsubsection{Removal of TypeScript}

In the original dataset, the first commit for \ts was recorded on
2003-03-21, several years before the language was created. Upon inspection,
we found that the file extension .\code{ts} is used for XML files containing
human language translations. Out of \initialNumTSProjects projects labeled
as \ts, only \realTSProjNum contained \ts. This reduced the number of
commits from \initialNumTSCommits to an even smaller \realTSCommitsNum.
Unfortunately, the three largest remaining projects (\code{\small
  typescript-node-definitions}, \code{\small Definitely\-Typed}, and the
deprecated \code{\small tsd}) contained only declarations and no code.  They
accounted for \pct{\ratioOfTypeDefTSCommits} of the remaining \ts commits.
Given the small size of the remaining corpus, we removed it from
consideration as it is not clear that we have sufficient data to draw useful
conclusions.  

To understand the origin of the classification error, we checked the tool
mentioned in the FSE paper, \gh
Linguist\footnote{\url{https://github.com/github/linguist}}. At the time of
the original study, that version of Linguist incorrectly classified
translation files as \ts. This was fixed on December 6th, 2014. This may
explain why the number of \ts projects decreased between the FSE and CACM
papers.

\subsubsection{Accounting for C++ and C}

Further investigation revealed that the input data only included \cpp
commits to files with the .\code{cpp} extension.  Yet, \cpp compilers allow
many extensions: .\code{C}, .\code{cc}, .\code{CPP}, .\code{c++},
.\code{cp}, and .\code{cxx}.  Moreover, the dataset contained no commits to
.\code{h} header files.  However, these files regularly contain executable
code such as inline functions in \c and templates in \cpp.  We could not
repair this without getting additional data and writing a tool to label the
commits in the same way as the authors did.  We checked \gh Linguist to
explain the missing files; but as of 2014, it was able to recognize header
files and all \cpp extensions.  

The only correction we applied was to delete the \code{V8} project. While V8
is written mostly in \cpp, its commits in the dataset are mostly in \js
(Fig.~\ref{v8} gives the number of commits per language in the dataset for
the V8 project).  Manual inspection revealed that \js commits were
regression test cases for errors in the missing \cpp code. Including them
would artificially increase the number of \js errors.  The original authors
may have noticed a discrepancy as they removed V8 from RQ3.

\begin{wrapfigure}{r}{3cm}
\footnotesize\vspace{2mm}
\begin{tabular}{@{}r|l@{}}
 & Commits\\
C     &  \vCCommits\\
C++&\vCppCommits\\
Python&\vPythonCommits\\
JavaScript&\vJavascriptCommits\\\hline
\end{tabular}
\caption{V8 commits.}\label{v8}
\end{wrapfigure}

\begin{figure}[!t]
\centering\includegraphics[width=.65\columnwidth]{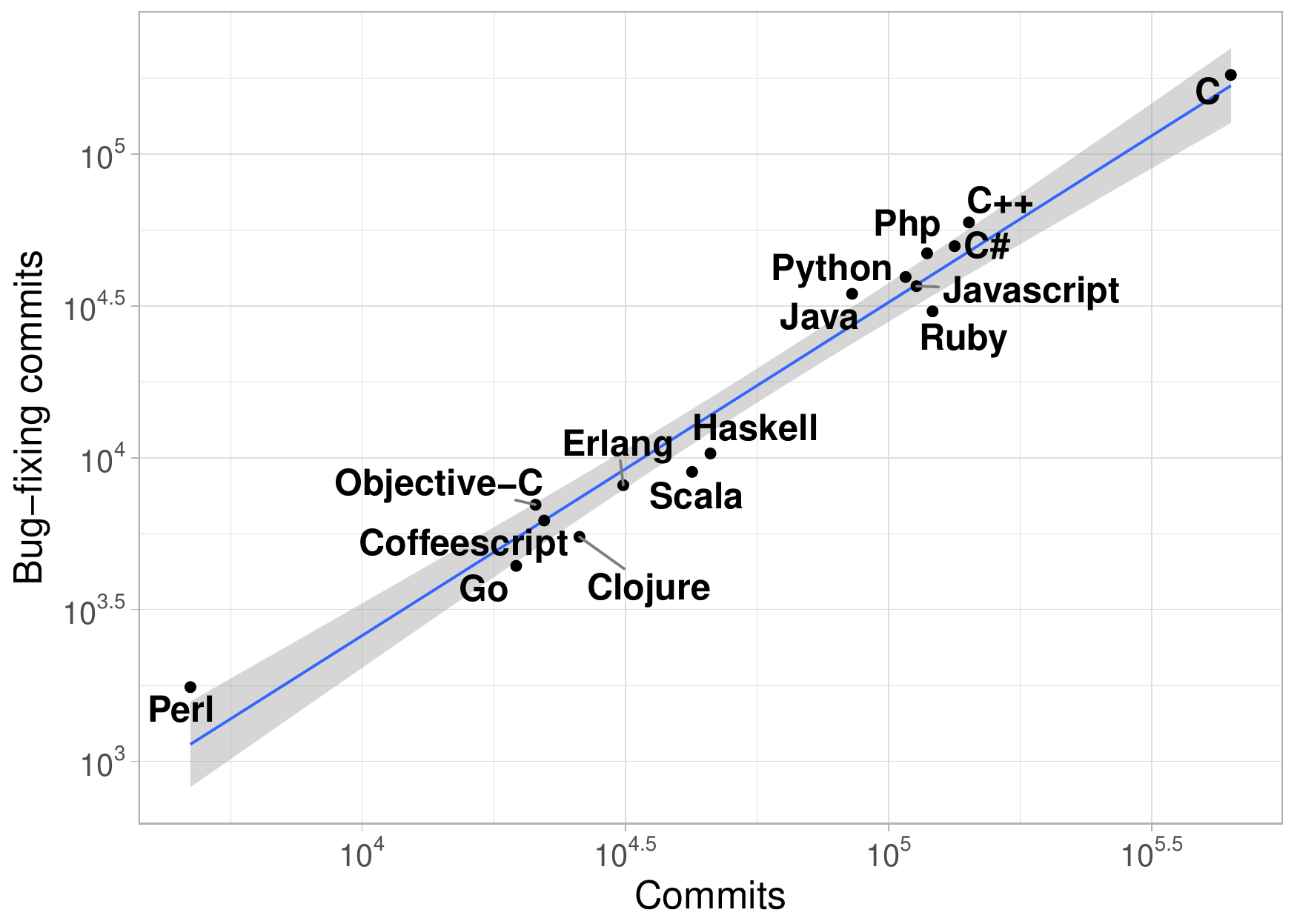}
\vspace{-1mm}
\caption{Commits and bug-fixing commits  after the cleaning.}\label{com}
\vspace{-3mm}\end{figure}

At the end of the data cleaning steps, the dataset had
\finalNumberOfProjectsIncluded projects, \finalSlocMio million lines of
code, and \finalNumShaMio million commits---of which \finalNumberOfBugFixes
were labeled as bug-fixing commits, written by \finalNumberAuthors thousand
authors.  Overall, our cleaning reduced the corpus by
\pct{\ratioReducedShaRows}.  Fig.~\ref{com} shows the relationship between
commits and bug fixes in all of the languages after the cleaning. As one
would expect, the number of bug-fixing commits correlated to the number of
commits. The figure also shows that the majority of commits in the corpus
came from \c and \cpp. \perl is an outliers because most of its commits were
missing from the corpus.

\subsubsection{Labeling Accuracy}\label{falselabels}

A key reanalysis question for this case study is: {\it What is a bug-fixing
  commit?}  With the help of 10 independent developers employed in industry,
we compared the manual labels of randomly selected commits to those obtained
automatically in the FSE paper.  We selected a random subset of 400 commits
via the following protocol. First, randomly sample 20 projects. In these
projects, randomly sample 10 commits labeled as bug-fixing and 10 commits
not labeled as bug-fixing. Enlisting help from 10 independent developers
employed in industry, we divided these commits equally among the ten
experts.  Each commit was manually evaluated by 3 of the experts. Commits
with at least 2 bugfix votes were considered to be bug fixes.  The review
suggested a false positive rate of 36\%; i.e., 36\% of the commits that the
original study considered as bug-fixing were in fact not. The false negative
rate was 11\%.  Short of relabeling the entire dataset manually, there was
nothing we could do to improve the labeling accuracy. Therefore, we chose an
alternative route and took labeling inaccuracy into account as part of the
statistical modeling and analysis.

We give five examples of commits that were labeled as bug fixing in the FSE
paper but were deemed by developers not to be bug fixes, each line contains
the text of the commit, underlined emphasis is ours and indicates the likely
reason the commit was labeled as a bug fix, the URL points to the commit in
\gh:
\begin{itemize}
\item {\tt tabs to spaces formatting \underline{fix}es}.\\ {\small\sf\url{https://langstudy.page.link/gM7N}} 
\item {\tt better \underline{error} messages}.\\ {\small\sf \url{https://langstudy.page.link/XktS}}
\item {\tt Converted   CoreDataRecipes sample to MagicalRecordRecipes sample application}.\\
 {\small\sf \url{https://langstudy.page.link/iNhr}}
\item {\tt [core] Add  NI\underline{Error}.h/m}.\\{\small\sf\url{https://langstudy.page.link/n7Yf}}
\item {\tt Add lazyness to in\underline{fix} operators}.\\{\small\sf\url{https://langstudy.page.link/2qPk}}
\end{itemize}

Unanimous mislabelings (when all three developers agreed) constituted 54\%
of the false positives. Within these false positives, most of the
mislabeling arose because words that were synonymous with or related to bugs
(e.g., ``fix'' and ``error'') were found within substrings or matched
completely out of context. A meta-analysis of the false positives suggests
the following six categories:
\begin{enumerate}
\item {\it Substrings};
\item {\it Non-functional:} meaning-preserving refactoring, e.g. changes to
  variable names;
\item {\it Comments:} changes to comments, formatting, etc.;
\item {\it Feature:} feature enhancements;
\item {\it Mismatch}: keywords used in unambiguous non-bug context
  (e.g. ``this isn't a bug'');
\item {\it Hidden features:} new features with unclear commit messages.
\end{enumerate}
The original study clarified that its classification, which involved
identifying bugfixes by only searching for error-related keywords came
from~\cite{Mockus00}.  However, that work classified modification requests
with an iterative, multi-step process, which differentiates between six
different types of code changes by through multiple keywords. It is possible
that this process was not completed in the FSE publication. In our citation
analysis, we encountered several papers that follow the same methodology as
the FSE paper for labeling, it is likely that their results need to be
adjusted as well.

\subsection{Methods: Statistical Modeling}

The reanalysis uncovered several methodological weaknesses in the
statistical analyses of the original manuscript.

\subsubsection{Zero-sum contrasts}\label{sec:zerosum}
The original manuscript chose to code the programming languages with
weighted contrasts.  Such contrasts interpret the coefficients of the
Negative Binomial Regression as deviations of the log-expected number of
bug-fixing commits in a language from the average of the log-expected number
of bug-fixing commits {\it in the dataset}. Comparison to the dataset
average is sensitive to changes in the dataset composition, makes the
reference unstable, and compromises the interpretability of the
results. This is particularly important when the composition of the dataset
is subject to uncertainty, as discussed in
Sec.~\ref{sec:methods:dataprocess} above.  A more common choice is to code
factors such as programming languages with zero-sum
contrasts~\cite{knnl}. This coding interprets the parameters as the
deviations of the log-expected number of bug-fixing commits in a language
from the average of log-expected number of bug-fixing commits {\it between
  the languages}. It is more appropriate for this investigation.

\subsubsection{Multiplicity of hypothesis testing}\label{sec:bonferroni}

A common mistake in data-driven software engineering is to fail to account
for multiple hypothesis testing~\cite{se-errors}. When simultaneously
testing multiple hypotheses, some p-values can fall in the significance
range by random chance. This is certainly true for Negative Binomial
Regression, when we simultaneously test 16 hypotheses of coefficients
associated with 16 programming languages being 0~\cite{knnl}. Comparing 16
independent p-values to a significance cutoff of, say, 0.05 in absence of
the associations implies the family-wise error rate (i.e., the probability
of at least one false positive association) FWER=$1-(1-0.05)^{16}=0.56$. The
simplest approach to control FWER is the method of Bonferroni, which
compares the p-values to the significance cutoff divided by the number of
hypotheses. Therefore, with this approach, we viewed the parameters as
``statistically significant" only if their p-values were below
$0.01/16=.000625$.

The FWER criterion is often viewed as overly conservative. An alternative
criterion is False Discovery Rate (FDR), which allows an average
pre-specified proportion of false positives in the list of ``statistically
significant" tests. For comparison, we also adjusted the p-values to control
the FDR using the method of Benjamini and Hochberg~\cite{fdr}.  An adjusted
p-value cutoff of, say, 0.05 implies an average 5\% of false positives in
the ``statistically significant" list.

As we will show next, for our dataset, both of these techniques agree in
that they decrease the number of statistically significant associations
between languages and defects by one (\ruby is not significant when we
adjust for multiple hypothesis testing).

\subsubsection{Statistical significance versus practical significance}

The FSE paper focused on statistical significance of the regression
coefficients.  This is quite narrow, in that the p-values are largely driven
by the number of observations in the dataset~\cite{halsey}. Small p-values
do not necessarily imply practically important
associations~\cite{lazar,misinterp}. In contrast, \emph{practical
  significance} can be assessed by examining model-based {\it prediction
  intervals}~\cite{knnl}, which predict future commits. Prediction intervals
are similar to confidence intervals in reflecting model-based
uncertainty. They are different from confidence intervals in that they
characterize the plausible range of values of the future individual data
points (as opposed to their mean). In this case study, we contrasted
confidence intervals and prediction intervals derived for individual
languages from the Negative Binomial Regression. As above, we used the
method of Bonferroni to adjust the confidence levels for the multiplicity of
languages.

\subsubsection{Accounting for uncertainty}

The FSE analyses assumed that the counts of bug-fixing commits had no
error. Yet, labeling of commits is subject to uncertainty, the heuristic
used to label commits is has many false positives, which must be factored
into the results. A relatively simple approach to achieve this relies on
parameter estimation by a statistical procedure called the
bootstrap~\cite{knnl}. We implemented the bootstrap with the following three
steps. First, we sampled with replacement the projects (and their
attributes) to create resampled datasets of the same size. Second, the
number of bug-fixing commits $\mathrm{bcommits}^*_i$ of project $i$ in the
resampled dataset was generated as the following random variable:

\vspace{-5mm}{\small\begin{eqnarray*}
\mathrm{bcommits}^*_i \sim \mathrm{Binom}(\mathrm{size}=\mathrm{bcommits}_i,\ \mathrm{prob=1-FP}) +\ \mathrm{Binom}(\mathrm{size}=(\mathrm{commits}_i-\mathrm{bcommits}_i),\ \mathrm{prob=FN}) 
\end{eqnarray*}}\vspace{-5mm}

\noindent where FP=36\% and FN=11\% (Section~\ref{sec:methods:dataprocess}).
Finally, we analyzed the resampled dataset with the Negative Binomial
Regression. The three steps were repeated 100,000 times to create the
histograms of estimates of each regression coefficients. Applying the
Bonferroni correction, the parameter was viewed as statistically significant
if 0.01/16th and (1-0.01)/16th quantiles of the histograms did not include  0.

\subsection{Results}

Table~\ref{langtable}(b-e) summarizes the re-analysis results. The impact of
the data cleaning, without multiple hypothesis testing, is illustrated by
column (b). Grey cells indicate disagreement with the conclusion of the
original work. As can be seen, the p-values for \c, \objc, \js, \ts, \php,
and \python all fall outside of the ``significant'' range of values, even
without the multiplicity adjustment. Thus, 6 of the original 11 claims are
discarded at this stage.  Column~(c) illustrate the impact of correction
for multiple hypothesis testing. Controlling the FDR increased the p-values
slightly, but did not invalidate additional claims. However, FDR comes at
the expense of more potential false positive associations. Using the
Bonferroni adjustment, does not change the outcome.  In both cases, the
p-value for one additional language, \ruby, lost its significance.

\begin{table*}[h]\begin{minipage}[t]{\textwidth}{\centering
\caption{Negative Binomial Regression for Languages}
\renewcommand{\G}{\cellcolor{gray!25}}
\begingroup\small
\scalebox{0.9}{
\begin{tabular}{@{}r||ll||ll|ll||ll||ll@{}}
  \hline
  \rule{0pt}{3ex} & \multicolumn{2}{c||}{\normalsize Original Authors}  & \multicolumn{8}{c}{\normalsize Reanalysis}\\[1mm] & \multicolumn{2}{c||}{(a) FSE~\cite{ray14}} & \multicolumn{2}{c|}{(b) cleaned data} & \multicolumn{2}{c||}{(c) pV adjusted} & \multicolumn{2}{c||}{(d) zero-sum} & \multicolumn{2}{c}{(e) bootstrap}\\ & Coef & P-val & Coef & P-val & FDR & Bonf & Coef & Bonf & Coef & sig.\\ \hline
Intercept & -1.93 & <0.001 & \cellcolor{white} -1.93 & \cellcolor{white} <0.001 & \cellcolor{white} -- & \cellcolor{white} -- & \cellcolor{white} -1.96 & \cellcolor{white} -- & \cellcolor{white} -1.79 & \cellcolor{white} * \\ 
  log commits & ~2.26 & <0.001 & \cellcolor{white} ~0.94 & \cellcolor{white} <0.001 & \cellcolor{white} -- & \cellcolor{white} -- & \cellcolor{white} ~0.94 & \cellcolor{white} -- & \cellcolor{white} ~0.96 & \cellcolor{white} * \\ 
  log age & ~0.11 & <0.01 & \cellcolor{white} ~0.05 & \cellcolor{white} <0.01 & \cellcolor{white} -- & \cellcolor{white} -- & \cellcolor{white} ~0.05 & \cellcolor{white} -- & \cellcolor{white} ~0.03 & \cellcolor{white}   \\ 
  log size & ~0.05 & <0.05 & \cellcolor{white} ~0.04 & \cellcolor{white} <0.05 & \cellcolor{white} -- & \cellcolor{white} -- & \cellcolor{white} ~0.04 & \cellcolor{white} -- & \cellcolor{white} ~0.03 & \cellcolor{white} * \\ 
  log devs & ~0.16 & <0.001 & \cellcolor{white} ~0.09 & \cellcolor{white} <0.001 & \cellcolor{white} -- & \cellcolor{white} -- & \cellcolor{white} ~0.09 & \cellcolor{white} -- & \cellcolor{white} ~0.05 & \cellcolor{white} * \\ 
   \hline
\hline
C & ~0.15 & <0.001 & \cellcolor{gray!25} ~0.11 & \cellcolor{gray!25} \hphantom{<}0.007 & \cellcolor{gray!25} \hphantom{<}0.017 & \cellcolor{gray!25} \hphantom{<}0.118 & \cellcolor{gray!25} ~0.14 & \cellcolor{gray!25} \hphantom{<}0.017 & \cellcolor{gray!25} ~0.08 & \cellcolor{gray!25}   \\ 
  C++ & ~0.23 & <0.001 & \cellcolor{white} ~0.23 & \cellcolor{white} <0.001 & \cellcolor{white} <0.01 & \cellcolor{white} <0.01 & \cellcolor{white} ~0.26 & \cellcolor{white} <0.01 & \cellcolor{white} ~0.16 & \cellcolor{white} * \\ 
  C\# & ~0.03 & -- & \cellcolor{white} -0.01 & \cellcolor{white} \hphantom{<}0.85 & \cellcolor{white} \hphantom{<}0.85 & \cellcolor{white} \hphantom{<}1 & \cellcolor{white} ~0.02 & \cellcolor{white} \hphantom{<}1 & \cellcolor{white} ~0 & \cellcolor{white}   \\ 
  Objective-C & ~0.18 & <0.001 & \cellcolor{gray!25} ~0.14 & \cellcolor{gray!25} \hphantom{<}0.005 & \cellcolor{gray!25} \hphantom{<}0.013 & \cellcolor{gray!25} \hphantom{<}0.079 & \cellcolor{gray!25} ~0.17 & \cellcolor{gray!25} \hphantom{<}0.011 & \cellcolor{gray!25} ~0.1 & \cellcolor{gray!25}   \\ 
  Go & -0.08 & -- & \cellcolor{white} -0.1 & \cellcolor{white} \hphantom{<}0.098 & \cellcolor{white} \hphantom{<}0.157 & \cellcolor{white} \hphantom{<}1 & \cellcolor{white} -0.07 & \cellcolor{white} \hphantom{<}1 & \cellcolor{white} -0.04 & \cellcolor{white}   \\ 
  Java & -0.01 & -- & \cellcolor{white} -0.06 & \cellcolor{white} \hphantom{<}0.199 & \cellcolor{white} \hphantom{<}0.289 & \cellcolor{white} \hphantom{<}1 & \cellcolor{white} -0.03 & \cellcolor{white} \hphantom{<}1 & \cellcolor{white} -0.02 & \cellcolor{white}   \\ 
  Coffeescript & -0.07 & -- & \cellcolor{white} ~0.06 & \cellcolor{white} \hphantom{<}0.261 & \cellcolor{white} \hphantom{<}0.322 & \cellcolor{white} \hphantom{<}1 & \cellcolor{white} ~0.09 & \cellcolor{white} \hphantom{<}1 & \cellcolor{white} ~0.04 & \cellcolor{white}   \\ 
  Javascript & ~0.06 & <0.01 & \cellcolor{gray!25} ~0.03 & \cellcolor{gray!25} \hphantom{<}0.219 & \cellcolor{gray!25} \hphantom{<}0.292 & \cellcolor{gray!25} \hphantom{<}1 & \cellcolor{gray!25} ~0.06 & \cellcolor{gray!25} \hphantom{<}0.719 & \cellcolor{gray!25} ~0.03 & \cellcolor{gray!25}   \\ 
  Typescript & -0.43 & <0.001 & \cellcolor{gray!25} -- & \cellcolor{gray!25} -- & \cellcolor{gray!25} -- & \cellcolor{gray!25} -- & \cellcolor{gray!25} -- & \cellcolor{gray!25} -- & \cellcolor{gray!25} -- & \cellcolor{gray!25} -- \\ 
  Ruby & -0.15 & <0.05 & \cellcolor{white} -0.15 & \cellcolor{white} <0.05 & \cellcolor{white} <0.01 & \cellcolor{gray!25} \hphantom{<}0.017 & \cellcolor{gray!25} -0.12 & \cellcolor{gray!25} \hphantom{<}0.134 & \cellcolor{white} -0.08 & \cellcolor{white} * \\ 
  Php & ~0.15 & <0.001 & \cellcolor{gray!25} ~0.1 & \cellcolor{gray!25} \hphantom{<}0.039 & \cellcolor{gray!25} \hphantom{<}0.075 & \cellcolor{gray!25} \hphantom{<}0.629 & \cellcolor{gray!25} ~0.13 & \cellcolor{gray!25} \hphantom{<}0.122 & \cellcolor{gray!25} ~0.07 & \cellcolor{gray!25}   \\ 
  Python & ~0.1 & <0.01 & \cellcolor{gray!25} ~0.08 & \cellcolor{gray!25} \hphantom{<}0.042 & \cellcolor{gray!25} \hphantom{<}0.075 & \cellcolor{gray!25} \hphantom{<}0.673 & \cellcolor{gray!25} ~0.1 & \cellcolor{gray!25} \hphantom{<}0.109 & \cellcolor{gray!25} ~0.06 & \cellcolor{gray!25}   \\ 
  Perl & -0.15 & -- & \cellcolor{white} -0.08 & \cellcolor{white} \hphantom{<}0.366 & \cellcolor{white} \hphantom{<}0.419 & \cellcolor{white} \hphantom{<}1 & \cellcolor{white} -0.05 & \cellcolor{white} \hphantom{<}1 & \cellcolor{white} ~0 & \cellcolor{white}   \\ 
  Clojure & -0.29 & <0.001 & \cellcolor{white} -0.31 & \cellcolor{white} <0.001 & \cellcolor{white} <0.01 & \cellcolor{white} <0.01 & \cellcolor{white} -0.28 & \cellcolor{white} <0.01 & \cellcolor{white} -0.15 & \cellcolor{white} * \\ 
  Erlang & ~0 & -- & \cellcolor{white} -0.02 & \cellcolor{white} \hphantom{<}0.687 & \cellcolor{white} \hphantom{<}0.733 & \cellcolor{white} \hphantom{<}1 & \cellcolor{white} ~0.01 & \cellcolor{white} \hphantom{<}1 & \cellcolor{white} -0.01 & \cellcolor{white}   \\ 
  Haskell & -0.23 & <0.001 & \cellcolor{white} -0.23 & \cellcolor{white} <0.001 & \cellcolor{white} <0.01 & \cellcolor{white} <0.01 & \cellcolor{white} -0.2 & \cellcolor{white} <0.01 & \cellcolor{white} -0.12 & \cellcolor{white} * \\ 
  Scala & -0.28 & <0.001 & \cellcolor{white} -0.25 & \cellcolor{white} <0.001 & \cellcolor{white} <0.01 & \cellcolor{white} <0.01 & \cellcolor{white} -0.22 & \cellcolor{white} <0.01 & \cellcolor{gray!25} -0.13 & \cellcolor{gray!25}   \\ 
   \hline
\end{tabular}
}
\endgroup
\label{langtable}}
\end{minipage}\end{table*}

Table~\ref{langtable}, column (d) illustrates the impact of coding the
programming languages in the model with zero-sum contrasts. As can be seen,
this did not qualitatively change the conclusions.  Table~\ref{langtable}(e)
summarizes the average estimates of coefficients across the bootstrap
repetitions, and their standard errors. It shows that accounting for the
additional uncertainty further shrunk the estimates closer to 0. In
addition, \scala is now out of the statistically significant set.

\begin{figure*}[!h]\begin{center}\begin{tabular}{c}
\hspace{-1.5cm}\includegraphics[width=1.2\textwidth]{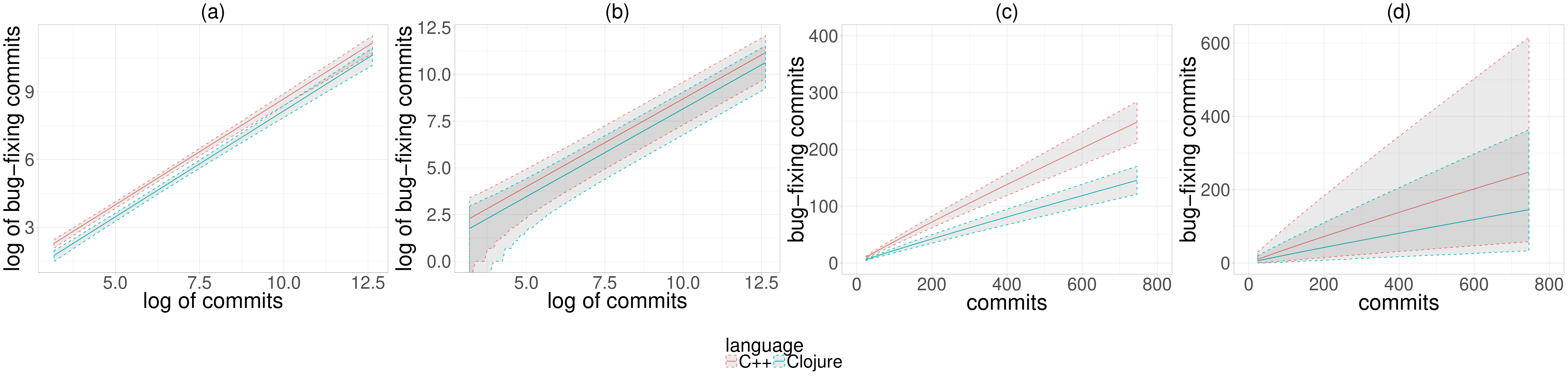}
\end{tabular}\end{center}\vspace{-6mm}
\caption{Predictions of bug-fixing commits as function of commits by models
  in Table~\ref{langtable}(c-d) for {C++} (most bugs) and {Clojure} (least
  bugs). (a) (1-0.01/16\%) confidence intervals for expected values on
  log-log scale. (b) Prediction intervals for a future number of bug-fixing
  commits, represented by $0.01/16$ and $1-0.01/16$ quantiles of the NB
  distributions with expected values in (a). (c)--(d): translation of the
  confidence and prediction intervals to the original
  scale.}\label{prediction}
\end{figure*}

\paragraph{Prediction intervals}
Even though some of the coefficients may be viewed as statistically
significantly different from 0, they may or may not be practically
significant. We illustrate this in Fig.~\ref{prediction}. The panels of the
figure plot model-based predictions of the number of bug-fixing commits as
function of commits for two extreme cases: \cpp (most bugs) commits) and
\clojure (least bugs).  Age, size, and number of developers were fixed to
the median values in the revised dataset. Fig.~\ref{prediction}(a) plots
model-based confidence intervals of the {\it expected values}, i.e. the
estimated average numbers of bug-fixing commits in the underlying population
of commits, on the log-log scale considered by the model. The differences
between the averages were consistently small. Fig.~\ref{prediction}(b)
displays the model-based {\it prediction intervals}, which consider
individual observations rather than averages, and characterize the plausible
future values of projects' bug-fixing commits. As can be seen, the
prediction intervals substantially overlap, indicating that, despite their
statistical significance, the practical difference in the future numbers of
bug-fixing commits is small. Fig.~\ref{prediction}(c)-(d) translate the
confidence and the intervals on the original scale and make the same point.

\subsection{Outcome}

The reanalysis failed to validate most of the claims of~\cite{ray14}.  As
Table~\ref{langtable}(d-f) shows, the multiple steps of data cleaning and
improved statistical modeling have invalidated the significance of 7 out of
11 languages. Even when the associations are statistically significant,
their practical significance is small.

\section{Follow up work}\label{threats}

We now list several issues that may further endanger
the validity of the causal conclusions of the original manuscript.

\subsection{Regression Tests}
Tests are relatively common in large projects. We discovered
\pct{\ratioTestsFiles} tests (\testFilesCommitted files) by matching file
names to the regular expression ``\code{*(Test|test)*}''. We sampled 100 of
these files randomly and verified that every one indeed contained 
regression tests. Tests are regularly modified to adapt to changes in API,
to include new checks, etc. Their commits may or may not be relevant, as
bugs in tests may be very different from bugs in normal code. Furthermore,
counting tests could lead to double counting bugs (that is, the bug fix and
the test could end up being two commits). Overall, more study is required to
understand how to treat tests when analyzing large scale repositories.

\subsection{Distribution of Labeling Errors}

Given the inaccuracy of automated bug labeling techniques, it is quite
possible that a significant portion of the bugs being analyzed are not
bug at all. We have shown how to accommodate for that uncertainty, but
our correction assumed a somewhat uniform distribution of labeling
errors across languages and projects.

Of course, here is no guarantee that labeling errors have a uniform
distribution.  Error rates may be influenced by practices such as using a
template for commits. For instance, if a project used the word \code{issue}
in their commit template, then automated tools would classify all commits
from that project as being bugs. To take a concrete example, consider the
\code{Design\-Patterns\-PHP} project, it has 80\% false positives, while
more structured projects such as \code{tengine} have only 10\% false
positives. Often, the indicative factor was as mundane as the wording used
in commit messages. The \code{gocode} project, the project with the most
false negatives, at 40\%, ``closes'' its issues instead of ``fixing'' them.
Mitigation would require manual inspection of commit messages and sometimes
even of the source code. In our experience, professional programmers can
make this determination in, on average, 2 minutes.  Unfortunately, this
would translate to 23 person-months to label the entire corpus.

\subsection{Project selection} 

Using \gh stars to select projects is fraught with perils as the 18 variants
of \code{bitcoin} included in the study attest.  Projects should be
representative of the language they are written in.  The
\code{PHP\-Design\-Patterns} is an educational compendium of code snippets,
it is quite likely that is does represent actual \php code in the wild.  The
\code{Definitely\-Typed} \ts project is a popular list of type signatures
with no runnable code; it has bugs, but they are mistakes in the types
assigned to function arguments and not programming errors.  Random sampling
of \gh projects is not an appropriate methodology either. \gh has large
numbers of duplicate and partially duplicated projects~\cite{oopsla17a} and
too many throwaway projects for this to yield the intended result.  To
mitigate this threat, researchers must develop a methodology for selecting
projects that represent the population of interest. For relatively small
numbers of projects, less than 1,000, as in the FSE paper, it is conceivable
to curate them manually. Larger studies will need automated techniques.

\subsection{Project provenance} 

\gh public projects tend to be written by volunteers working in open source
rather than by programmers working in industry. The work on many of these
projects is likely to be done by individuals (or collections of individuals)
rather than by close knit teams. If this is the case, this may impact the
likelihood of any commit being a bug fix. One could imagine commercial
software being developed according to more rigorous software engineering
standards.  To mitigate for this threat, one should add commercial projects
to the corpus and check if they have different defect characteristics.
If this is not possible, then one should qualify the claims by describing
the characteristics of the developer population.

\subsection{Application domain} 

Some tasks, such as system programming, are inherently more challenging and
error prone than others. Thus, it is likely that the source code of an
operating system may have different characteristics in terms of error than
that of a game designed to run in a browser. Also, due to non-functional
requirements, the developers of an operating system may be constrained in
their choice of languages (typically unmanaged system languages).  The
results reported in the FSE paper suggest that this intuition is wrong.  We
wonder if the choice of domains and the assignment of projects to domains
could be an issue. A closer look may yield interesting observations.

\begin{figure}[!h]\includegraphics[width=.5\columnwidth]{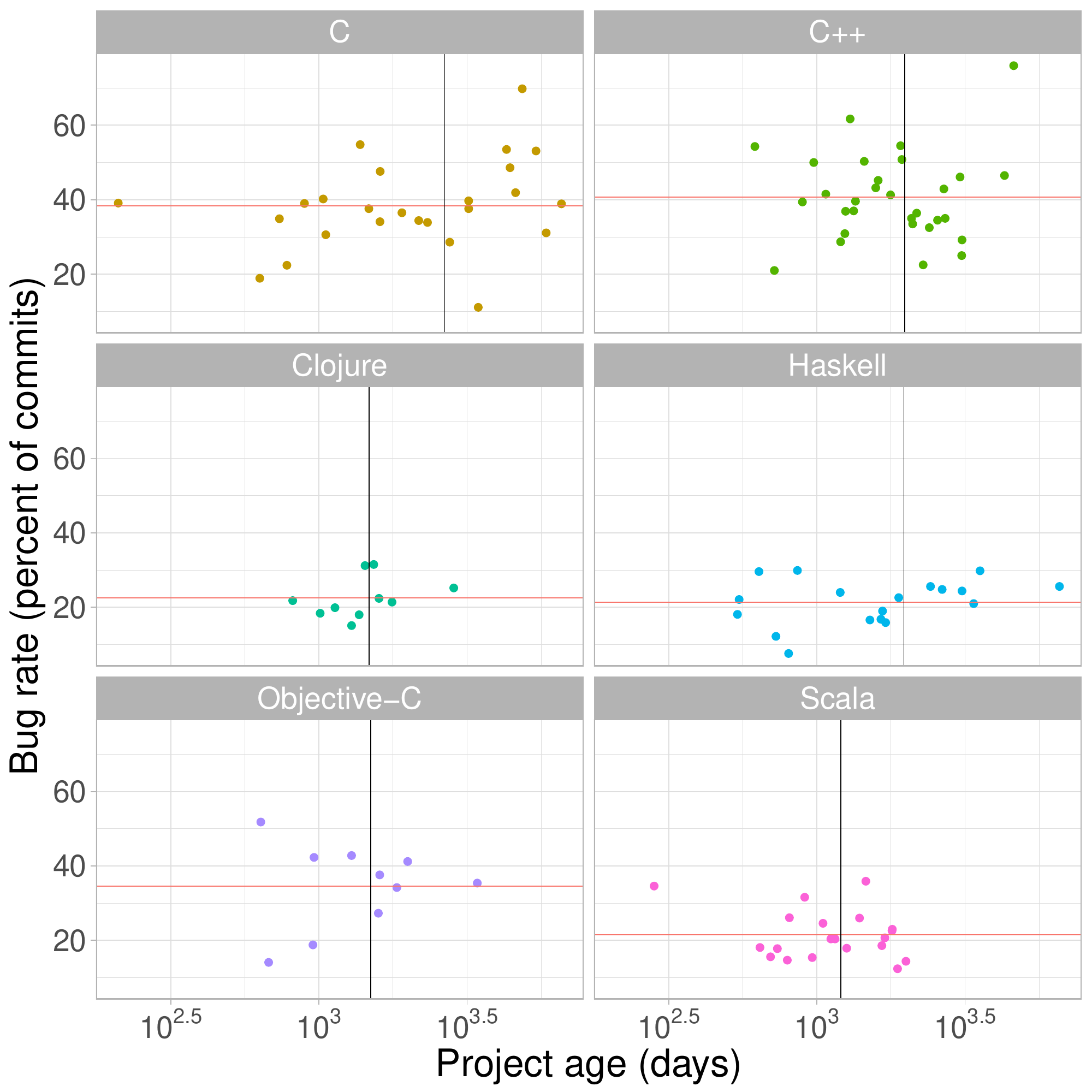}
\vspace{-3mm}
\caption{Bug rate vs. project age. Lines indicate means of project age
  (x-axis) and bug rate (y-axis).}\label{age}
\end{figure}

\subsection{Uncontrolled influences} 

\begin{figure}[!h]
\includegraphics[width=.6\columnwidth]{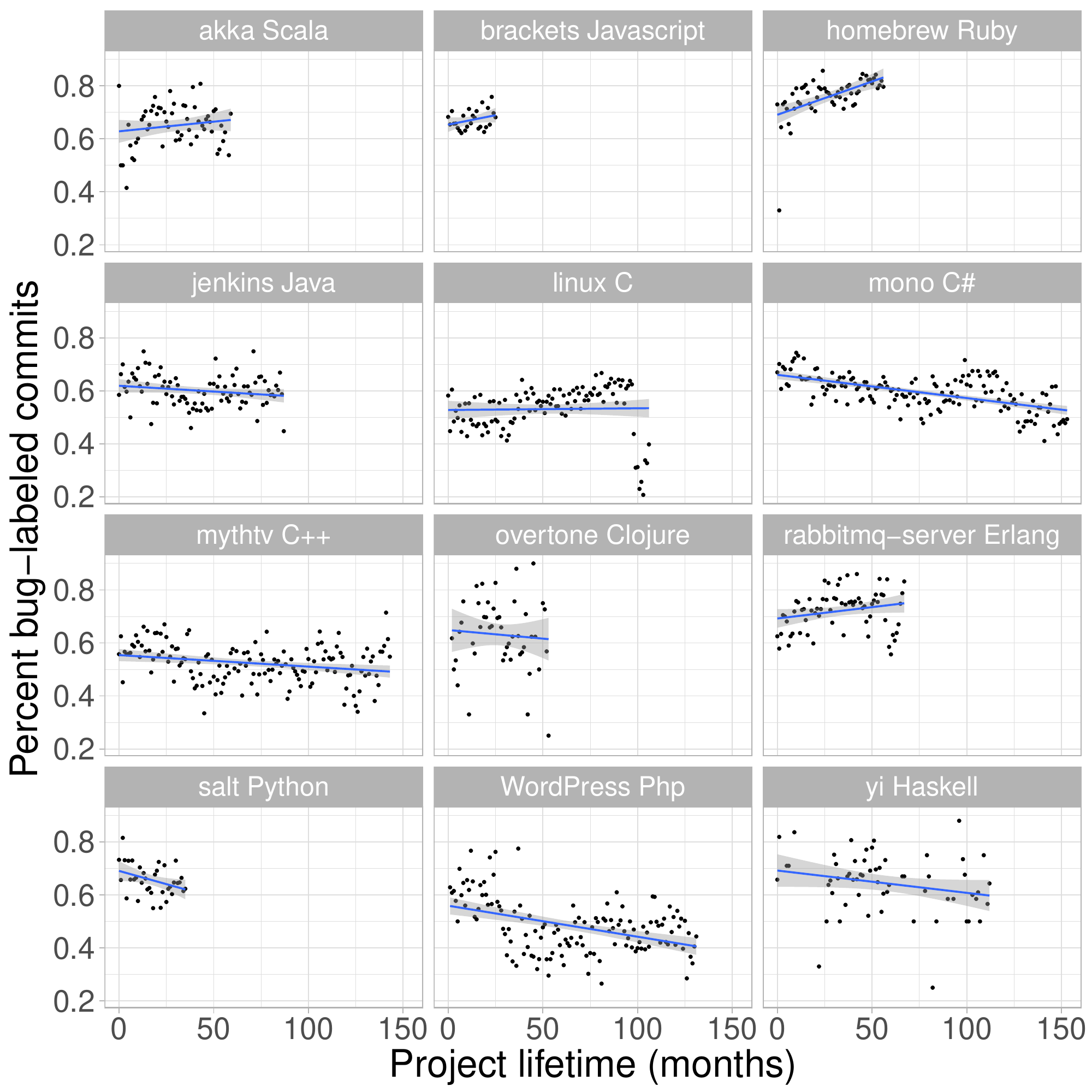}
\vspace{-3mm}
\caption{Monthly avg. bug rate over lifetime. Points are
 \% of bug-labeled commits, aggregated over months.}\label{evolution}
\end{figure}

Additional sources of bias and confounding should be appropriately
controlled. The bug rate (number of bug-fixing commits divided by total
commits) in a project can be influenced by the project's culture, the age of
commits, or the individual developers working on it.  Consider
Fig.~\ref{age}, which shows that project ages are not uniformly distributed:
some languages have been in widespread use longer than others.  The relation
between age and its bug rate is subtle. It needs to be studied, and the age
should be factored into the selection of projects for inclusion in the
study.  Fig.~\ref{evolution} illustrates the evolution of the bug rate over
time for 12 large projects written in various languages. While the projects
have different ages, there are clear trends. Generally, bug rates decrease
over time. Thus, older projects may have a smaller ratio of bugs, making the
language they are written in appear less error-prone. Lastly, the FSE paper
did not control for developers influencing multiple projects. While there
are over 45K developers, 10\% of these developers are responsible for 50\%
of the commits.  Furthermore, the mean number of projects that a developer
commits to is 1.2. This result indicates that projects are not
independent. To mitigate those threats, further study is needed to
understand the impact of these and other potential biases, and to design
experiments that take them into account.

\subsection{Relevance to the RQ} 
The FSE paper argues that programming language features are, in part,
responsible for bugs. Clearly, this only applies to a certain class of
programming errors: those that rely on language features.  Bugs related to
application logic or characteristics of the problem domain are less likely
to be affected by the programming language. For example, setting the wrong
TCP port on a network connection is not a language-related bug, whereas
passing an argument of the wrong data type may be. It is eminently possible
that the majority of bugs are in fact not affected by language features. To
mitigate this threat, one would need to develop a new classification of
bugs. It is unclear what attributes of a bug would be used for this purpose
and quite unlikely that the process could be conducted without manual
inspection of the source code.

\section{Best Practices}\label{best}

The lessons from this work mirror challenges of reproducible data science.
While these lessons are not novel, they may be worth repeating.

\subsection{Automate, document, and share} 
The first lesson touches upon the process of collecting, managing, and
interpreting data.  Real-world problems are complex, and produce rich,
nuanced, and noisy datasets. Analysis pipelines must be carefully engineered
to avoid corruption, errors, and unwarranted interpretations. This turned
out to be a major hurdle for the FSE paper. Uncovering these issues on our
side was a substantial effort (approximately 5 person-months). 

Data science pipelines are often complex: they use multiple languages, and
perform sophisticated transformations of the data to eliminate invalid
inputs and format the data for analysis. For instance, this paper relies on
a combination of \js, {\sf\small R}, {\sf\small shell}, and {\sf\small
  Makefiles}. The {\sf\small R} code contains over 130 transformation
operations over the input table.  Such pipelines can contain subtle
errors---one of the downsides of statistical languages is that they almost
always yield a value. Publications often do not have the space to fully
describe all the statistical steps undertaken. For instance, the FSE paper
did not explain the computation of weights for NBR in sufficient detail for
reproduction.  Access to the code was key to understanding.  However, even
with the source code, we were not able to repeat the FSE results---the code
had suffered from bit rot and did not run correctly on the data at hand.
The only way forward is to ensure that all data analysis studies be (a)
automated, (b) documented, and (c) shared.  Automation is crucial to ensure
repetition and that, given a change in the data, all graphs and results can
be regenerated. Documentation helps understanding the analysis. A pile of
inscrutable code has little value.

\subsection{Apply domain knowledge} 
Work in this space requires expertise in a number of disparate areas.
Domain knowledge is critical when examining and understanding
projects. Domain experts would have immediately taken issue with V8 and
bitcoin. Similarly, the classification of \scala as a purely functional
language or of \objc as a manually managed language would have been red
flags. Finally, given the subtleties of Git, researchers familiar with that
system would likely have counseled against simply throwing away merges. We
recognize the challenge of developing expertise in all relevant technologies
and concepts. At a minimum, domain experts should be enlisted to vet claims.

\subsection{Grep considered harmful} 

Simple analysis techniques may be too blunt to provide useful answers.  This
problem was compounded by the fact that the search for keywords did not look
for words and instead captured substrings wholly unrelated to software
defects. When the accuracy of classification is as low as 36\%, it becomes
difficult to argue that results with small effect sizes are meaningful as
they may be indistinguishable from noise.  If such classification techniques
are to be employed, then a careful \emph{post hoc} validation by hand should
be conducted by domain experts.

\subsection{Sanitize and validate} Real-world data is messy.
Much of the effort in this reproduction was invested in gaining a thorough
understanding of the dataset, finding oddities and surprising features in
it, and then sanitizing the dataset to only include clean and tidy
data~\cite{hadley}. For every flaw that we uncovered in the original study
and documented here, we developed many more hypotheses that did not pan out.
The process can be thought of as detective work---looking for clues, trying
to guess possible culprits, and assembling proof.

\subsection{Be wary of  p-values}
Our last advice touches upon data modeling, and model-based
conclusions. Complicated problems require complicated statistical analyses,
which in turn may fail for complicated reasons. A narrow focus on
statistical significance can undermine results. These issues are well
understood by the statistical community, and are summarized in a recent
statement of the American Statistical Association~\cite{lazar}. The
statement makes points such as ``{\it Scientific conclusions should not be
  based only on whether a p-value passes a specific threshold}'' and ``{\it
  A p-value, or statistical significance, does not measure the importance of
  a result.}''  The underlying context, such as domain knowledge, data
quality, and the intended use of the results, are key for the validity of
the results.

\section{Conclusion}

The Ray et al.\ work aimed to provide evidence for one of the fundamental
assumptions in programming language research, which is that language design
matters. For decades, paper after paper was published based on this very
assumption, but the assumption itself still has not been validated. The
attention the FSE and CACM papers received, including our reproduction
study, directly follows from the community's desire for answers.

Unfortunately, our work has identified numerous problems in the FSE study
that invalidated its key result.  Our intent is not to blame, performing
statistical analysis of programming languages based on large-scale code
repositories is hard.  We spent over 6 months simply to recreate and
validate each step of the original paper.  Given the importance of the
questions addressed by the original paper, we believe it was time well
spent. Our contribution not only sets the record straight, but more
importantly, provides thorough analysis and discussion of the pitfalls
associated with statistical analysis of large code bases. Our study should
lend support both to authors of similar papers in the future, as well as to
reviewers of such work.

Finally, we would like to reiterate the need for automated and reproducible
studies. While statistical analysis combined with large data corpora is
a powerful tool that may answer even the hardest research questions, the
work involved in such studies---and therefore the possibility of errors---is
enormous. It is only through careful re-validation of such studies that
the broader community may gain trust in these results and get better insight
into the problems and solutions associated with such studies. 

\paragraph{Acknowledgments} 
We thank Baishakhi Ray and Vladimir Filkov for sharing the data and code of
their FSE paper. Had they not preserved the original files and part of their
code, reproduction would have been more challenging.  We thank Derek Jones,
Shiram Krishnamurthi, Ryan Culppeper, Artem Pelenitsyn for helpful comments

\newpage
\bibliographystyle{acmart/ACM-Reference-Format}\bibliography{main}
\end{document}